\documentstyle[natbib,amsfonts,epsfig,12pt,here]{article}
\def\be{\begin{equation}}
\def\ee{\end{equation}}

\begin{document}
\begin{titlepage}

\centerline{{\Large \bf Fluctuation Analysis of the Far-Infrared Background -}}
\vskip 0.2cm
\centerline{{\Large \bf Information from the Confusion}}
\vskip 0.5cm
\centerline{\large  Yasmin Friedmann$^{1,2}$ and Fran${\c{} c}$ois Bouchet$^{3}$}
\vskip 0.4cm
\centerline{$^1$ Racah Institute, The Hebrew University, Jerusalem, 91904, Israel}
\vskip 0.1cm
\centerline{$^2$Departement de Physique Theorique, University of Geneva} 
\vskip 0.1cm
\centerline{24, quai E. Ansermet, 1211 Geneve 4, Switzerland}
\vskip 0.1cm
\centerline{$^3$ IAP, 98bis, Bd Arago 75014 Paris, France}
\vskip 1.5cm

\begin{abstract}

We are investigating to what extent one can use a P(D) analysis to
extract number counts of unclustered sources from maps of the far infrared
background.  Currently available such maps, and those expected to
emerge in near future are dominated by confusion
noise due to poor resolution. We simulate background maps with an
underlying two slope model for N(S) and we find that in an experiment
of FIRBACK type we can extract the high flux slope with an error of
few percent while other parameters are not so well constrained. We
find, however, that in a SIRTF type  experiment all parameters of
this N(S) model can be extracted with errors of only few percent.
\end{abstract}
\end{titlepage}

\section{Introduction}

Analysis of spatial fluctuations in the level of background radiation 
has been used by radio \cite[]{sch74,condon74} and x-ray astronomers
\cite[]{bar&fab90} to gain information on number count-flux relation
below the detection limit.  

Recently \cite[]{lag&pug00}, fluctuations in the infrared background
were first detected in maps of the FIRBACK survey at wavelength of
170$\mu $m. 

Spatial resolution currently available at the far infra-red is of the
order of arcminutes. As a result observations at this wavelength are confusion
limited. This means that the dominant contribution to noise on sky
maps at these wavelengths comes not from detector or photon noise but from the
superposition of light originating from galaxies which are too close
on the sky to be resolved individually. It has been shown
\cite[]{firback1} that the energy coming from resolved sources on the
FIRBACK maps comprise only 10$\%$ of the  total energy while the rest
is due to the unresolved background radiation. This means that other
than fluctuation analysis ('P(D) analysis'), not much else can be done
to study the N(S) of the unresolved infrared sources at the far infra-red. 

This study investigates under which conditions P(D) analysis can
usefully constrain galaxy evolution scenarios. 
\cite{guider98} have introduced a semi-analytical model of
galaxy formation and evolution, and within this model suggested
several scenarios including different amounts of ultra luminous
infrared galaxies. For each of the scenarios they calculated, among other
things, faint galaxy counts. They show that at 175 $\mu $m the source
counts at fluxes 10-100 mJy are quite sensitive to 
the details of the galaxy evolution; therefore knowing N(S) to high
precision can help choosing between the different scenarios of their
galaxy evolution  models. Similarly \cite{takeuchi} show that at 175
$\mu $m the number counts at fluxes 10-100 mJy are very dependent on the
galaxy evolution models they are suggesting.

When using a simple power law parametrization for the source counts, 
errors of few percent in the parameters can give an
error of ~50$\%$ in the number counts at fluxes of few 10s of
mJys, i.e. one anticipates that such counts at these flux levels will
easily constrain the parameters. Additionally, at least in the above
mentioned families of models at this flux range, the N(S) 
due to the different models differ by an order of magnitude. This
justifies attempting to measure the N(S) parameters to high
precision down to 
few 10s mJy. Note that the flux range of few 10s mJy is far below the
detection limit of FIRBACK (180 mJy) and therefore, at the moment, can
be probed only via P(D) analysis.  
 
Additionnally one would also like to know how much information one can gain
about the number counts from specific confusion limited surveys like FIRBACK
or SIRTF using a P(D) analysis. This kind of study can be done
using a Fisher Matrix Analysis where one calculates the minimal
errors of extracted parameters given the experiment and a parameterized
theoretical model. 

The analysis we make here does not take into account clustering of
sources. Some clustering of the far infrared sources is of course
expected \cite[]{knox01,haiman00,scott99}, but its
amplitude is not yet determined at 175 $\mu $m. The small
area of the FIRBACK fields might not enable to accurately constrain
the source clustering but this situation might change with the SIRTF
observations which cover 
larger areas of the sky \cite[]{dole_sirtf}. The resolved sources on
the FIRBACK maps show a level of clustering consistent with zero. This
is probably due 
to the small number of resolved sources (Guiderdoni and Lagache, private
communication). Hence we are assuming, at this stage, that sources are
distributed poissonianly on the sky.  

We find that we can constrain the slope of the number counts of
sources with high fluxes ($\geq ~20$ mJy) at least as well as has
been done by extracting individual strong sources. Other parameters
(slope of the number counts at low fluxes, normalization and break
flux) are not as well constrained in the FIRBACK type of experiment.
However, in an experiment with smaller pixels and more of them
(e.g. SIRTF) we can extract all the parameters to within several percent.

We also found some degeneracies between the  different parameters and saw
that better experiment like SIRTF cannot resolve these degeneracies.

The paper is arranged as follows. We give an explanation of the nature
of confusion in sky surveys in section 2. In section 3 we describe how
we model the N(S) in order to extract it from the data. In section 4
we outline the method of analyzing sky maps in order to extract the
parameters of the model. We describe the implementation of the method
and the results of analyzing simulated skies in section 5.

\section{Confusion}

The spatial fluctuations in the level of background radiation due to spatial
distribution of  the discrete sources which contribute to the
background are called confusion noise. In the far infra-red
wavelengths the level of the confusion noise dominates over any photon
or instrumental detector noise existing in today's
instruments. Thus while one can reduce the level of instrumental
or photon noise by long integration times, the confusion remains a
strong characteristic of far-infrared observations.

The full calculation of the probability distribution of the fluctuations (P(D) for
short) is classical (see \cite[]{condon74,bar&fab90}) and given in the
appendix. Here we only give the final result: 
\begin{equation}
P(D)=\int_{-\infty}^{\infty}\Phi(\omega)e^{-2\pi i\omega D}d\omega
\end{equation}
where D is the deflection from the mean level of flux and
\begin{equation}
\Phi(\omega)=exp\left[A_{pix}\int_0^{\infty}dS\frac{dN}{dS}(S)\sum_i\left(e^{2\pi
      i\omega P_iS}-1\right)\right].
\end{equation}

The shape of the P(D) depends on several inputs. First is the 
differential N(S) relation, $\frac{dN}{dS}(S)$. This is the number of
sources per steradian with fluxes in [S,S+dS]. It also depends on the
shape of the beam, which is described by $P_i$. $P_i$ is the point
spread function of the telescope and $A_{pix}$ is the pixel size.. 
In general the P(D) will also depend on clustering of the sources if
it is strong enough \cite[]{barcons92} but as mensioned before, we are going to deal
with a source distribution which is not clustered but Poissonian.  

It was shown in \cite{sch74} that the width of the curve (the
1$\sigma$ of the noise) is of the order of the flux for which  there
is one source per beam. The very faint sources do not 
contribute at all to the shape of the curve, but only to the mean
level of the flux. This is because there are very many of them within
each beam and the change of their number from beam to beam is
relatively small and so does not contribute to the fluctuations. The
very strong sources contribute only to the tail of the distribution. 
Typically the flux where there is one source per beam is much lower
then the resolution limit. 

\section{LogN-LogS of infrared sources}

Since this work is motivated by the FIRBACK survey we will give in the
following a short description of the survey and of the
$\frac{dN}{dS}(S)$ found for the resolved sources. These details will
guide us when we construct simulations to check our method of deriving 
$\frac{dN}{dS}(S)$ from the observed P(D).

\subsection{The FIRBACK survey}

The FIRBACK \cite[]{firback1,firback2,firback3} is a deep survey of 4
square degrees of the sky at 170$\mu $m. The 4 degrees
were chosen in such a way that the foreground cirrus contamination was as small
as possible. Thus one can get information on the extra-galactic
radiation \cite[]{lag&puget00a}.
In the FIRBACK survey there were 106 sources detected above the
sensitivity limit of the experiment at 4$\sigma$, with fluxes between
180 mJy to 2400 mJy. The slope of the  LogN-LogS curve was measured by
\cite{firback3} to be $-3.3 \pm 0.6$ between 180 and 500 mJy.

\subsection{Modeling Source Counts}

In view of the above we will assume a broken power law for the
source-count model: the slope at low fluxes has to become shallower
than 3.0 or the flux per pixel will diverge. Another motivation for
this two slope model is the predictions coming from galaxy evolution
models discussed earlier. In all of the predictions the number counts
exhibit a relative flattening at low fluxes.  

Therefore we write $\frac{dN(S)}{dS}$  as follows:

\begin{equation}
\frac{dN(S)}{dS} = \left\{ \begin{array}{ll}
A_{norm}S^{-\gamma_1} & \textrm{for $S\ge S_{break}$}\\
A_{norm}S_{break}^{\gamma_2-\gamma_1}S^{-\gamma_2} & \textrm{for $S_{min}\le
S\le S_{break}$} \end{array} \right.
\label{model_ns}
\end{equation}

The parameters to be determined are the normalization, $A_{norm}$, 
the flux of the break in the power-law, $S_{break}$, and the two
slopes, $\gamma_1$ and $\gamma_2$. Since $S_{min}$ does not
change the shape of the distribution and only affects the mean flux,
and since we will be fitting for the shape of the distribution and not
for the mean flux, $S_{min}$ becomes irrelevant to the fitting
process. It comes into play only in order to fine tune the mean flux
to its value from data. In the
following we will refer to the four parameters commonly as $\theta$, and the
probability distribution of the deflection will be written as P(D;$\theta$).

\section{Method of  Analysis}
\subsection{Minimum $\chi^2$  Method for Binned Data}
Our data set is composed of several thousand measurements of incoming flux
received by 46$\times $46 arcsec$^2$ pixels which are pointed to
different directions in the sky. We will bin the data according to
flux. In this way we can compare the experimental distribution of the
fluctuations to a calculated P(D;$\theta$).  

Binning data may proceed in two ways. One way is such that the bins
are equal in length and the number of events vary from bin to bin. Or
one may bin the data such that there is the same number of events in 
each of the bins and the size of the bins changes accordingly. We use
the former method but we manually increase the bin size at the two
tails of the distribution where there are very few events, so as not
to have bins with zero events. We have thus 3 bins where there are
around 5 events per bin, out of a total of ~60 bins.  

A histogram is in fact a multinomial distribution. This is the
generalization of the binomial distribution to the case where it is
possible to have more then two outcomes for the experiment. In our case
the flux received by a pixel pointing in one of the directions in the
sky will be one result of the experiment, and the outcome might fall in any
one of the bins. Let's call $r_1,...r_n$ the $n$
possibilities for the outcome of the experiment (in our case $n$
different flux ranges from the minimal to the maximal flux received in
the map), and let $p_i$ be the probability for a pixel to fall in bin
$i$. The sum of all these 
probabilities is of course 1 since every pixel falls in one of the
bins. We assume that the different pixels are independent (we will
justify this assumption in section 5.2) . Then, after N
experiments of measuring the incoming flux (N pixels in the map), the
probability that the fluxes will distribute with  $r_1,...,r_n$ 
pixels falling in the bins will be given by   
\begin{equation}
P({r_i},N,{p_i})=\frac{N!}{r_1!r_2!...r_n!} p_1^{r_1} p_2^{r_2}...p_n^{r_n}
\end{equation}
Some important properties of this distribution are that $E(r_i)$, the
expectation value for the bin i, is given by  $E(r_i)=Np_i$ and
$V(r_i)$, the variance for bin i is given by
$V(r_i)=Np_i(1-p_i)$. When the number of experiments becomes 
large the multinomial distribution tends to the multi-normal 
distribution.  In the case when there are many bins and $p_i<<1$, the
variance tends to be the expectation value.  
Thus when we bin the fluxes we must take care to have enough bins such
that on the one hand $p_i<<1$ and on the other, there are enough
pixels per bin so that the distribution of the number of pixels per
bin will be close to Gaussian. Then the overall likelihood
of the data can be written as  follows:
\begin{equation}
L(r_1,....r_n;\theta)=Ce^{-\frac{1}{2}Q^2}
\end{equation}
where $Q^2$, the quadratic form, is
\begin{equation}
Q^2=\sum_{i=1}^n\sum_{j=1}^n c_{ij}\left(\frac{r_i-\mu _i}{\sigma_i}\right)\left(\frac{r_j-\mu _j}{\sigma_j}\right)
\end{equation}
$c_{ij}$ are the elements of the inverse covariance matrix C, given by
\begin{equation}
  \label{eq:cov_mat}
  C=E[(\stackrel{\rightarrow}{r}-\stackrel{\rightarrow}\mu )(\stackrel{\rightarrow}{r}-\stackrel{\rightarrow}\mu )^T] 
\end{equation}
$r_i$ is the number of pixels within flux bin i, $\mu _i$ is the 
expected number of pixels in the bin according to the model, and
$\sigma_i$ is the square root of the variance, in our case it is $\sqrt{\mu _i}$. 
The correlation $\rho_{i,j}$ between bins i,j is given by 
\begin{equation}
  \label{eq:cor_mat}
\rho_{i,j}=C_{i,j}/\sigma_i\sigma_j
\end{equation}
If we assume that there are no or only negligible correlations between
the errors in the bins then the covariance matrix will be 
almost diagonal. Therefore the quantity we need to minimize becomes   
\begin{equation}
Q^2=\sum_{i=1}^{N_{bins}}\left(\frac{r_i-\mu _i}{\sqrt{\mu _i}}\right)^2
\end{equation}
This method is the ``minimum chi-square method'' applied to histogram
fits. When there are many events in each bin then $Q^2$ has
asymptotically a $\chi^2$ distribution with [n-(number of
fitted parameters)] degrees of freedom and the method is equivalent to a
maximum likelihood estimation method. Therefore from now on we will
use the notation $\chi^2=Q^2$. We are assuming that our way of binning
allows $Q^2$ to behave close enough to $\chi^2$.

\subsection{Fisher Matrix Analysis}

As described before, we have several thousand measurements of the flux
S, and a model for N(S) which leads to a P(D, $\theta$) and we
estimate the parameters  $\theta$  using maximum likelihood method
with the binned data. There is a lower bound to the variance of an estimator, which is
related to the  Fisher Matrix, $F_{ij}$ 
\begin{equation}
F_{ij} = E\left[-\frac{\partial log L}{\partial \theta_i \partial \theta_j}\right]
\end{equation}
The Rao-Cramer-Frechet inequality states that for any unbiased
estimator, $\Delta \theta_i\ge(F^{-1})_{ii}^{1/2}$ where $\Delta
\theta_i$ is the 1$\sigma$ error of the parameter $\theta_i$. This
inequality was used by several authors 
\cite[e.g.]{weigh96,efst99,karh97} to assess how well may different
parameters be estimated in future experiments. 

In our case, where the the likelihood is multinormal, and the errors
depend on the parameters, the Fisher matrix becomes \cite[]{karh97}:
\be
F_{ij}=\frac{1}{2}\sum_{k=1}^{N_{bins}}\frac{\partial \mu _k}{\partial
  \theta_i}\frac{\partial \mu _k}{\partial
  \theta_j}\frac{1}{\mu _k}\left[2+\frac{1}{\mu _k}\right]\bigg|_{\theta_{ML}}
\label{fm}
\ee
As we can see, the Fisher matrix depends solely on the model and on
the estimated parameters, $\{\theta_{ML}\}$. 
We can now calculate the minimal variance for different experimental
setups, namely as a function of the number of pixels in the maps, or
as a function of the size of the pixel. In this way we can foresee how
well we could extract the 4 parameters from future experiments.

The inverse of the Fisher matrix is an estimate of the
covariance matrix, so by investigating F$^{-1}$ one can locate
degeneracies between parameters and see if they might be resolved 
in different experimental setups. 

\section{Results}

In this section we describe the analysis of simulated skies that we have
made in order to check theoretically the limits of our method using
the fisher matrix analysis.

\subsection{Simulations}

\begin{figure*}
\begin{center}
$\begin{array}{cc}
\multicolumn{1}{l}{} & \multicolumn{1}{l}{} \\ 
\includegraphics[width=7cm]{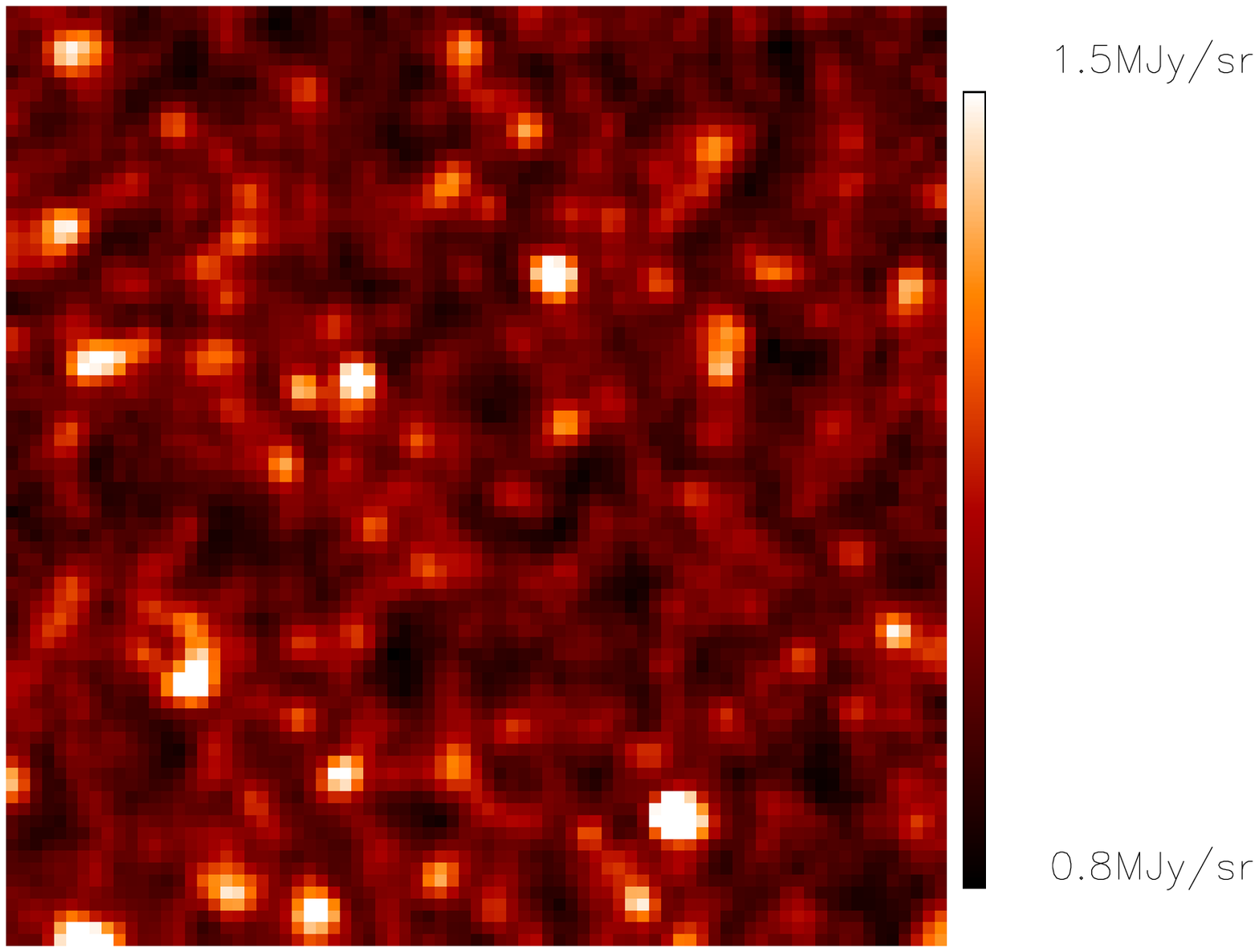} &
\includegraphics[width=7cm]{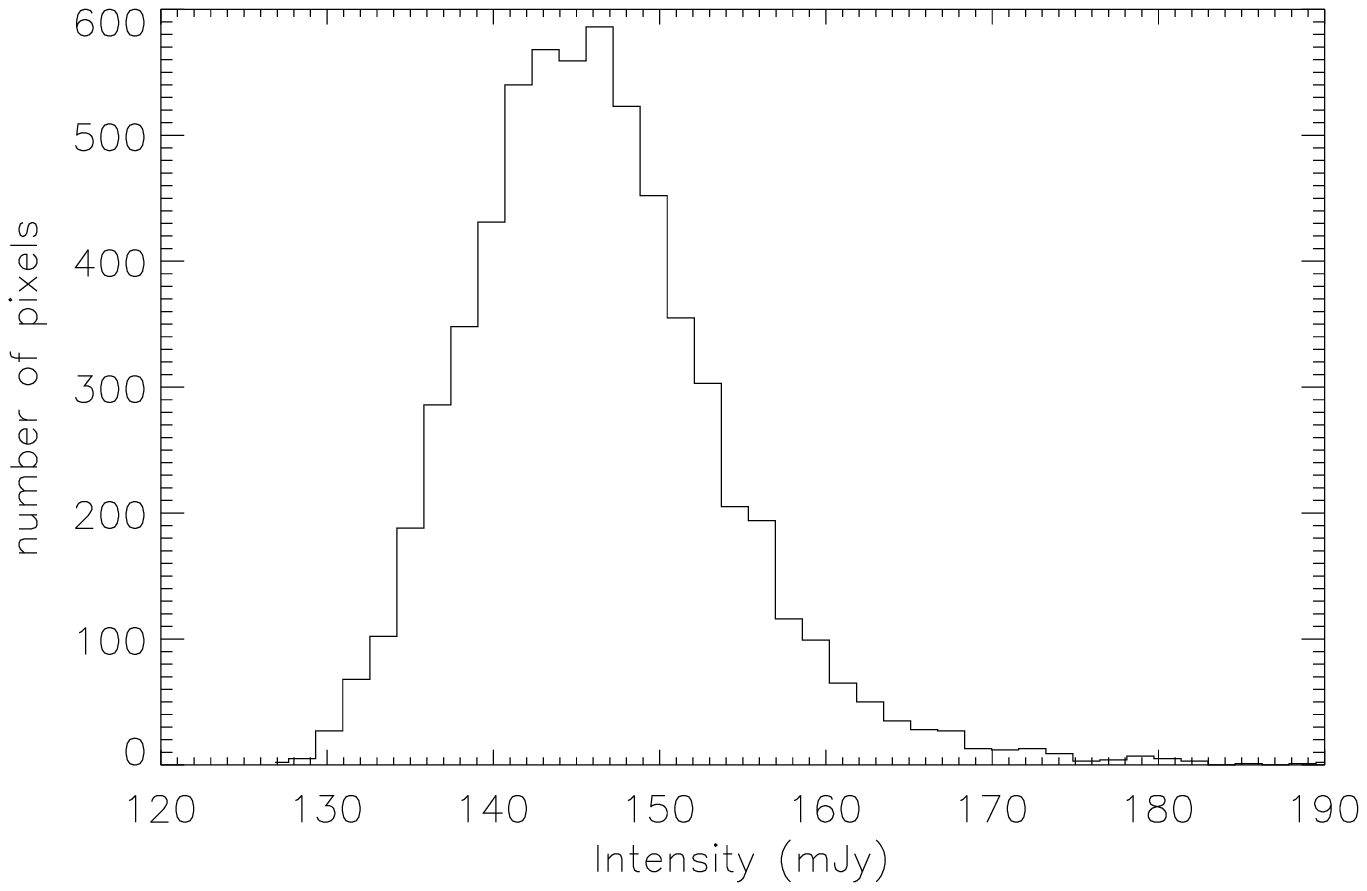} \\ 
\end{array}$
\end{center}
\caption{Left: 1 square degree of simulated sky, with pixels of 46$^2$
  arcsec$^2$. The sky contains some $2*10^8$ sources, described by a
  model  N(S) given in equation \ref{model_ns}, with parameters as
follows: $\gamma_1=3.3$, $\gamma_2=1.8$, $S_{break}=14$mJy and
$A_{norm}=0.18*10^{11}$. Right: The histogram of the simulated map showing
the fluctuations in the intensity. Note the skewed shape, with the
tail due to the strong sources.} 
\label{map}
\end{figure*}

We use mock images to check our algorithm. These images are
built in two steps. First we build a mock sky. This is a high
resolution projected image of a given 
spatial distribution of sources. It is constructed by distributing
point sources randomly on a $1.25 \times 1.25$ square degree field. Each 
source is assigned a flux such that the overall number counts are
consistent  with our predetermined two-slope N(S). We
chose to model N(S) such that the main traits of the FIRBACK maps are
realized, such as the number of sources above 180mJy and the mean
level of the background. Then we convolve it with the instrumental
effects in order to produce an image as close as possible to real
data. Finally we extract only the central 1 square degree as our map.  
A map produced this way is presented in figure \ref{map} (a), and its
histogram is shown in \ref{map}(b). The map includes some $2*10^8$
sources with fluxes ranging from 0.01 to 2000 mJy, and the parameters
chosen were $\gamma_1=3.3$, $\gamma_2=1.8$ and $S_{break}=14$mJy and
the normalization was 0.18*$10^{11}$.

\subsection{Effect of Different PSFs on Correlation Between Flux Bins}

The total extent of the PSF of the FIRBACK instrument and of the
upcoming SIRTF is 
bigger then the size of a pixel in these experiments. In FIRBACK the
Full Width at Half Maximum of the PSF is equal to the pixel size. For
SIRTF it is twice as wide \cite{dole_sirtf}.
In such cases we expect there to be some correlation between adjacent pixels on the
map. Given this, we want to quantify to what
extent these correlations may be manifested as correlations between the
different bins in the histogram. 

To this end we should measure the correlation coefficients of the
histogram given different sized PSFs. The different PSFs we used were
constructed from the original one by rebinning again and again. 
We quantify their extent with respect to the pixel size by defining
variable x which is the ratio of the FWHM to the pixel size. The x we
used are 2.7, 1.1, 0.5, 0.3 and 0.1.  
Following this we produced hundreds of sky maps based on the same
underlying N(S) relation. We convolved each with a FSF, produced a
histogram, and measured the sample correlation matrix.
The process was repeated for the different PSFs, and was done once
without convolving with any PSF but instead we rebinned the sky to the
size of the pixel, 46''$\times $ 46''.

\begin{figure*}
\begin{center}
\includegraphics[width=7.3cm]{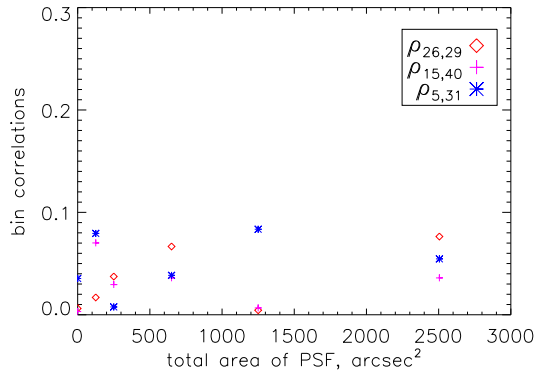}
\end{center}
\caption{Three randomly chosen elements of the correlation matrix,
  plotted as a function of the total coverage area of the PSF. All
  correlations are less then 10$\%$, and there is no clear trend of
  behavior as a function of the coverage.}
\label{cor_ij_psf}
\end{figure*}

If bins were not correlated at all, the correlation matrix would have 
been the identity matrix. We found that there is a negligible level of
correlations with any of the PSFs, mostly less than 10$\%$. The largest
correlation seen was between few neighboring bins for the largest
PSF, near the center of the histogram, at a level of ~20$\%$. There is
also no clear behavior of  $\rho_{ij}$ as a function of the total coverage
area of the PSF. This can be seen in figure $\ref{cor_ij_psf}$, where
3 randomly selected elements of the correlation matrix are plotted as
a function of the total coverage area of the PSF.
In light of this we will neglect the correlations between the bins 
and estimate the parameters as described in the next section. 

\subsection{Estimating Parameters}

\begin{figure*}
\begin{center}
\includegraphics[width=7cm]{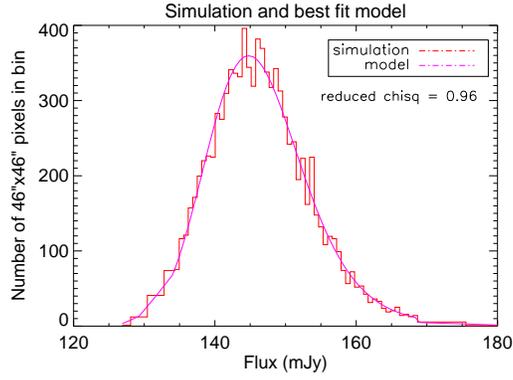}
\end{center}
\caption{Here we see again the histogram of the simulated map with the best fit model}
\label{mapandfit}
\end{figure*}

In order to find estimates for
the 4 parameters $A_{norm},S_{break},\gamma_1,\gamma_2$, we first
choose a small enough value for $S_{min}$. On the one hand it should
be smaller then reasonable values of $S_{break}$, and on the other
hand it should be big enough to avoid numerical problems of integration. 
We then grid a large enough part of parameter space around a
reasonable point found by trial and error. Each of these grid
points will serve as a starting point for the minimization
procedure. This way we have more chance of catching the lowest
minimum. The minimization procedure calculates the $\chi^2$ and uses
the derivatives of the model to go downhill in parameter space until
a lowest local $\chi^2$ is found. We looked at the estimated
parameters and their errors to see whether they are all situated in
the same part of the 
parameter space. We repeated the procedure with several different
binning to make sure that the results are not dependent on the binning.   
The best fit is shown in figure \ref{mapandfit}. The reduced $\chi^2$
for this model is 0.96. The errors of the estimated parameters will be
discussed below.

\begin{figure*}
\begin{center}
$\begin{array}{c@{\hspace{0.1in}}c}
\multicolumn{1}{l}{} & \multicolumn{1}{l}{} \\ 
\includegraphics[width=7cm]{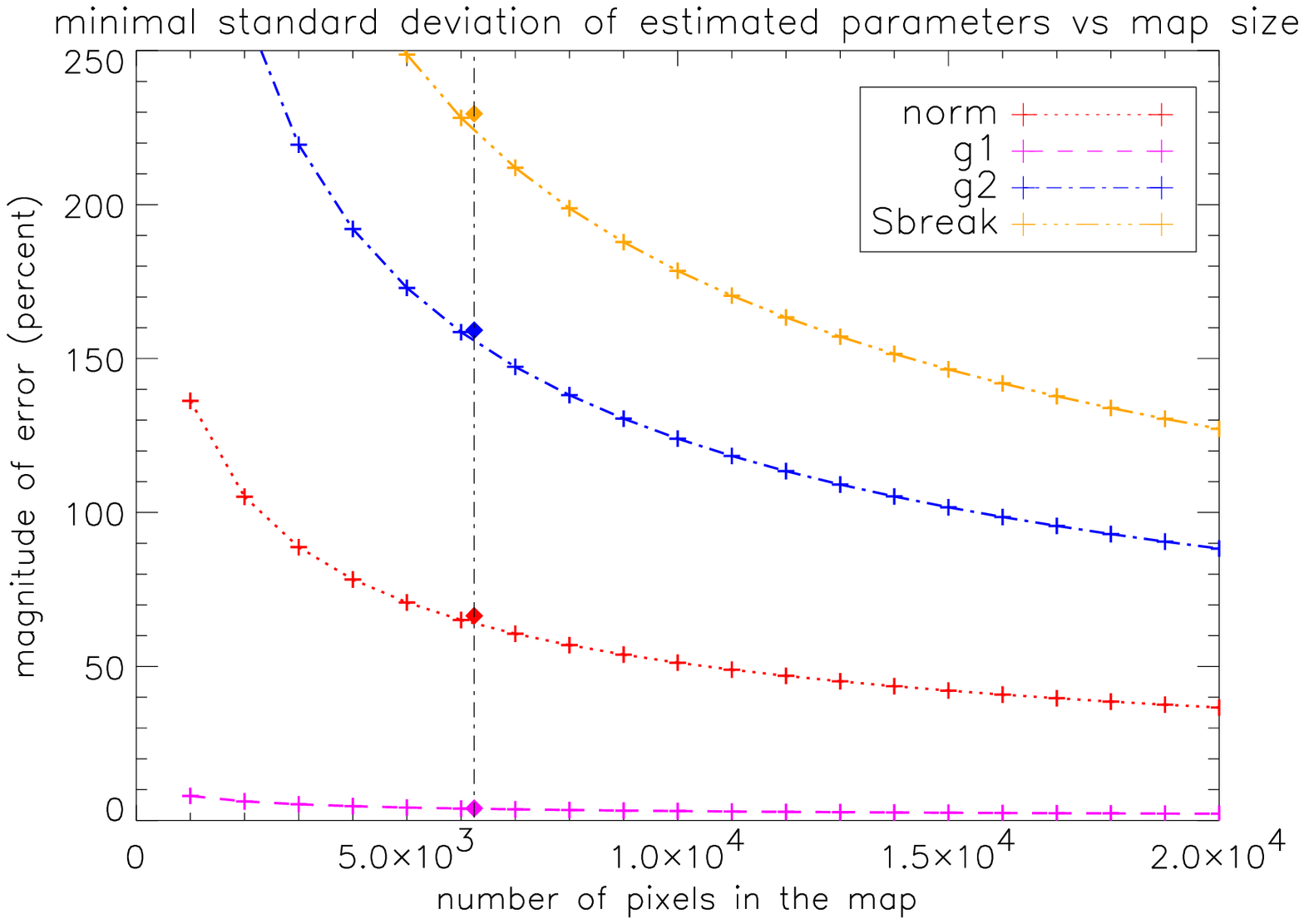} &
\includegraphics[width=7cm]{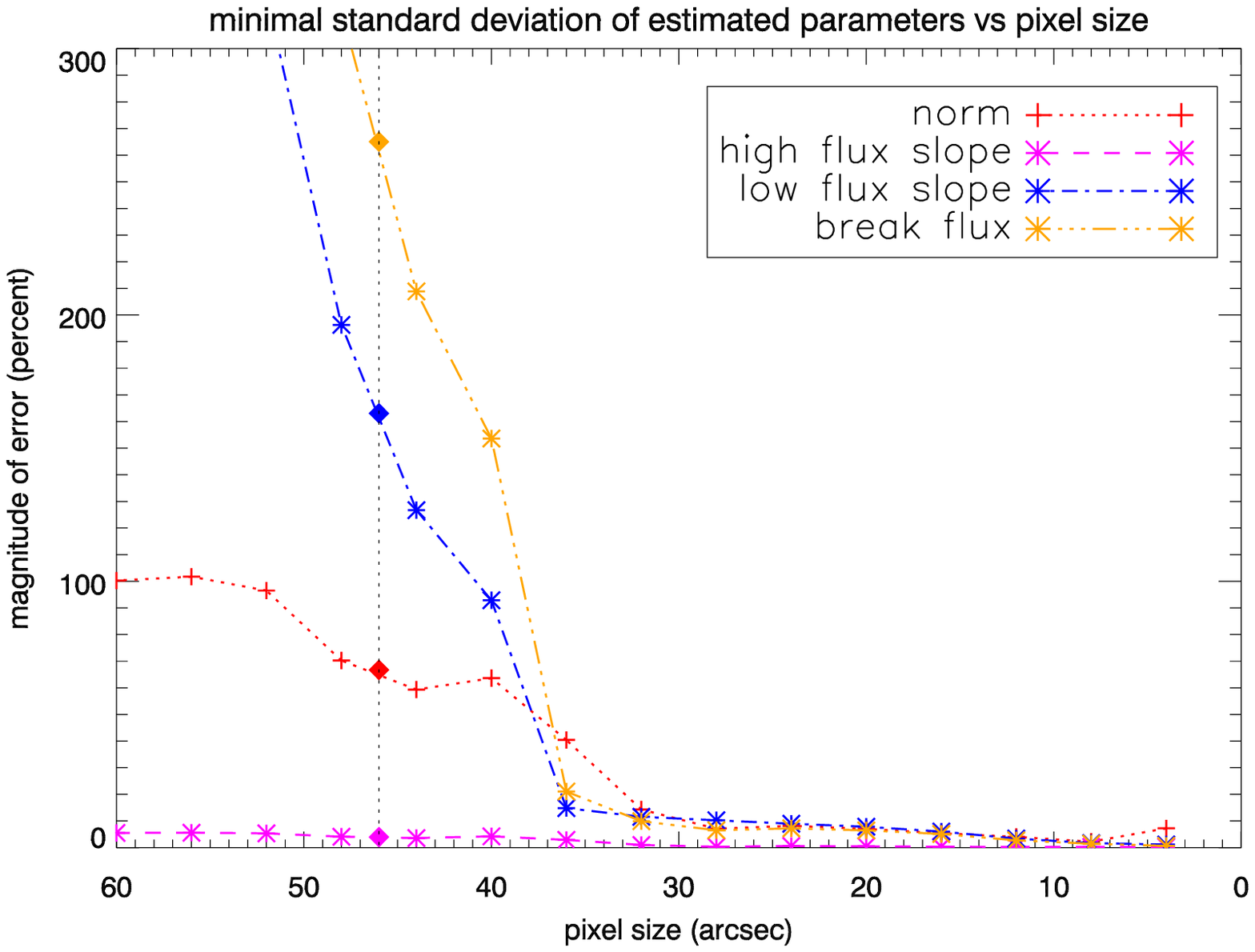} \\ 
\mbox{(a)} & \mbox{(b)}
\end{array}$
\end{center}
\caption{Shown are the errors on the estimated parameters (in percent)
  as a function of the number of pixels in
  the map (a) and as a function of the pixel size (b). The vertical
  line shows where the FIRBACK experiment is on these grafs.}
\label{mapsize}
\end{figure*}

Once we have a best fit set of parameters, we assume that it is
a good enough approximation of the real parameters. Then we can continue to
calculate the fisher matrix elements for different experimental
parameters.  We will calculate the minimal errors of the different parameters
as a function of pixel size and as a function of the number of pixels in the map.
In fig. \ref{mapsize}a one sees how the errors reduce when we add more
pixels to the map. In fact they reduce as (number of pixels)$^{-1}$ -
and this behavior is what we expect from the definition of the fisher
matrix, equation \ref{fm}. 
In  fig. \ref{mapsize}b one sees that the errors greatly reduce once we use
smaller pixel sizes in the experiment, with all errors at the level of only
few percent once we reach pixel size of 15 arcsec$^2$. This is
very encouraging because in the upcoming SIRTF experiment the pixels will be
of size $16\times 16$ arcsec$^2$. 

Another point to look at is the comparison between the errors on the
estimated parameters in our algorithm, and the minimal errors given
from the Fisher Matrix analysis. To this end we plotted, on both
\ref{mapsize}a and \ref{mapsize}b a vertical line indicating the
properties of the FIRBACK and our simulations. On top of this line we
put the parameter errors we got from the minimization procedure. It is
encouraging to see that these errors are only very slightly larger than the minimal
errors possible according to the Cramer-Rao-Frechet lower bound.

Another way to see the great improvement in the precision of the
estimation is when one looks at the 1$\sigma$ contours which enclose
68$\%$ of the joint distribution of several pairs  of estimators while
we marginalize over the other two parameters, as in fig. \ref{ellipses}.
The contours, independent of which plane we look at, encompass a
shrinking patch of the parameter space as we add pixels to the map or
use smaller pixels. We have plotted the original parameters as filled
hexagons on top of the ellioses. It is important to note, also, that except for the
normalization, the true parameters of the simulation are enclosed
within the 1$\sigma$ contours of the original experiment. 

\subsection{Parameter Degeneracies}

\begin{figure*}
\begin{center}
$\begin{array}{c@{\hspace{0.1in}}cc}
\multicolumn{1}{l}{} & \multicolumn{1}{l}{} & \multicolumn{1}{l}{}\\ 
\includegraphics[width=5.0cm]{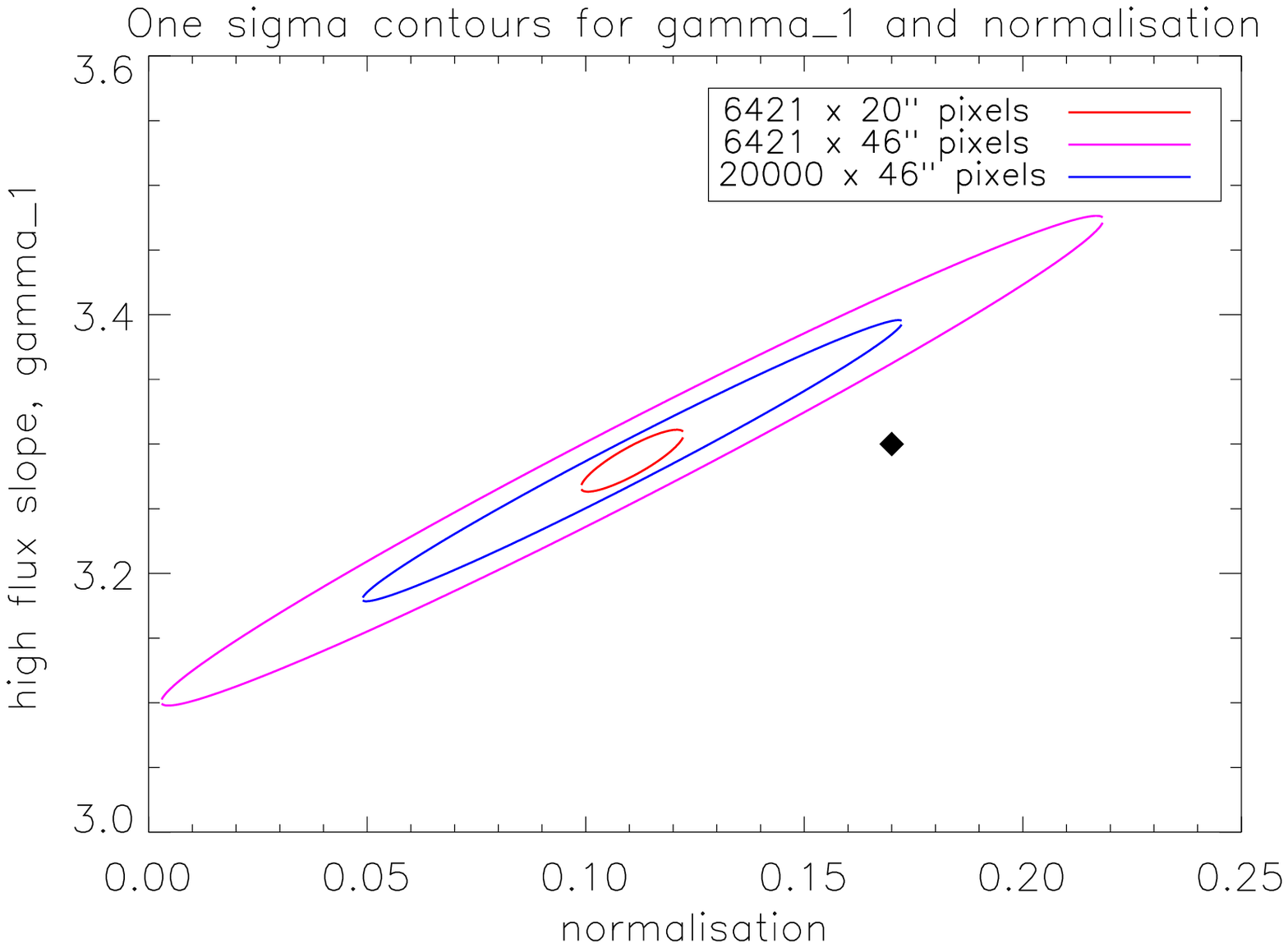} &
\includegraphics[width=5.0cm]{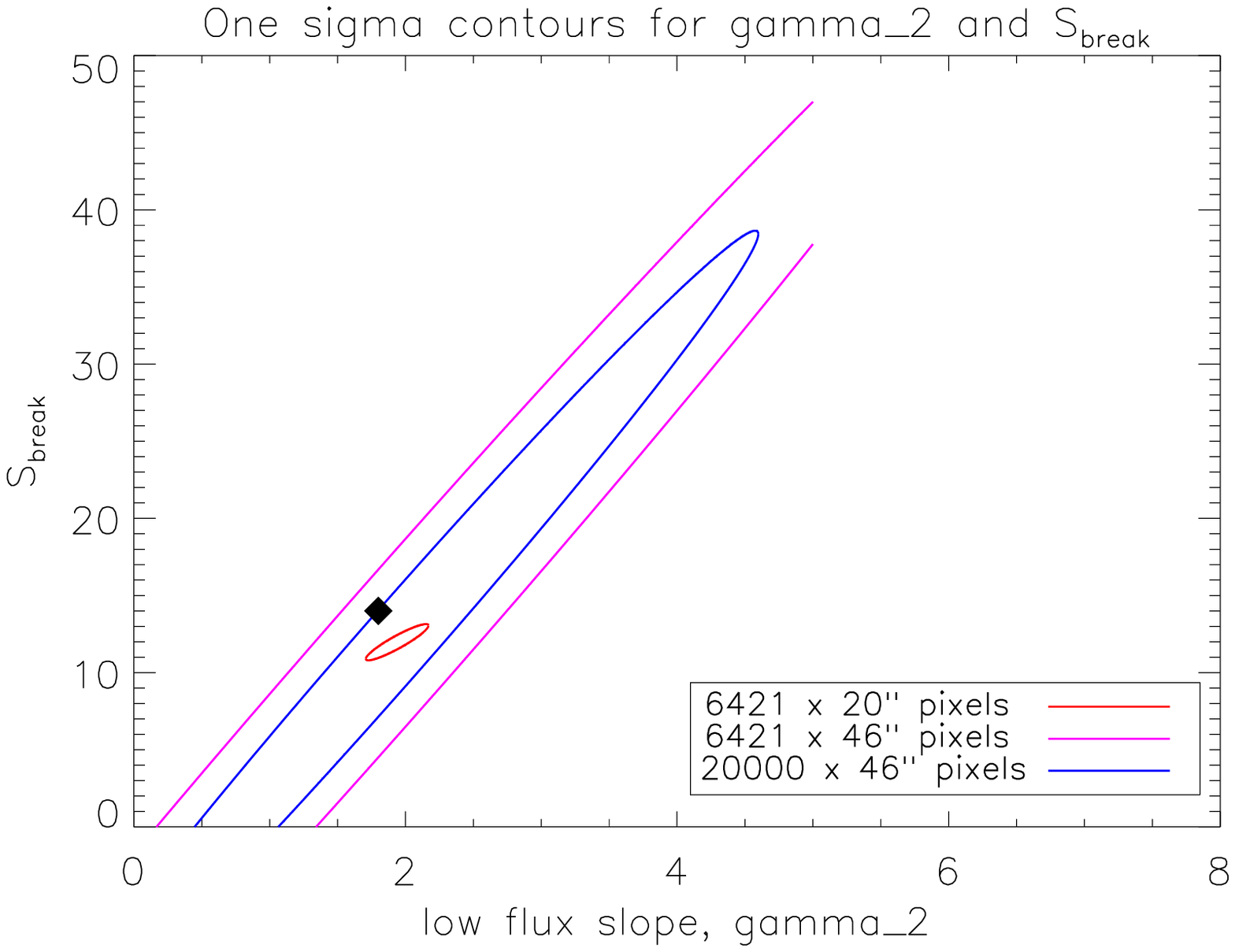} &
\includegraphics[width=5.0cm]{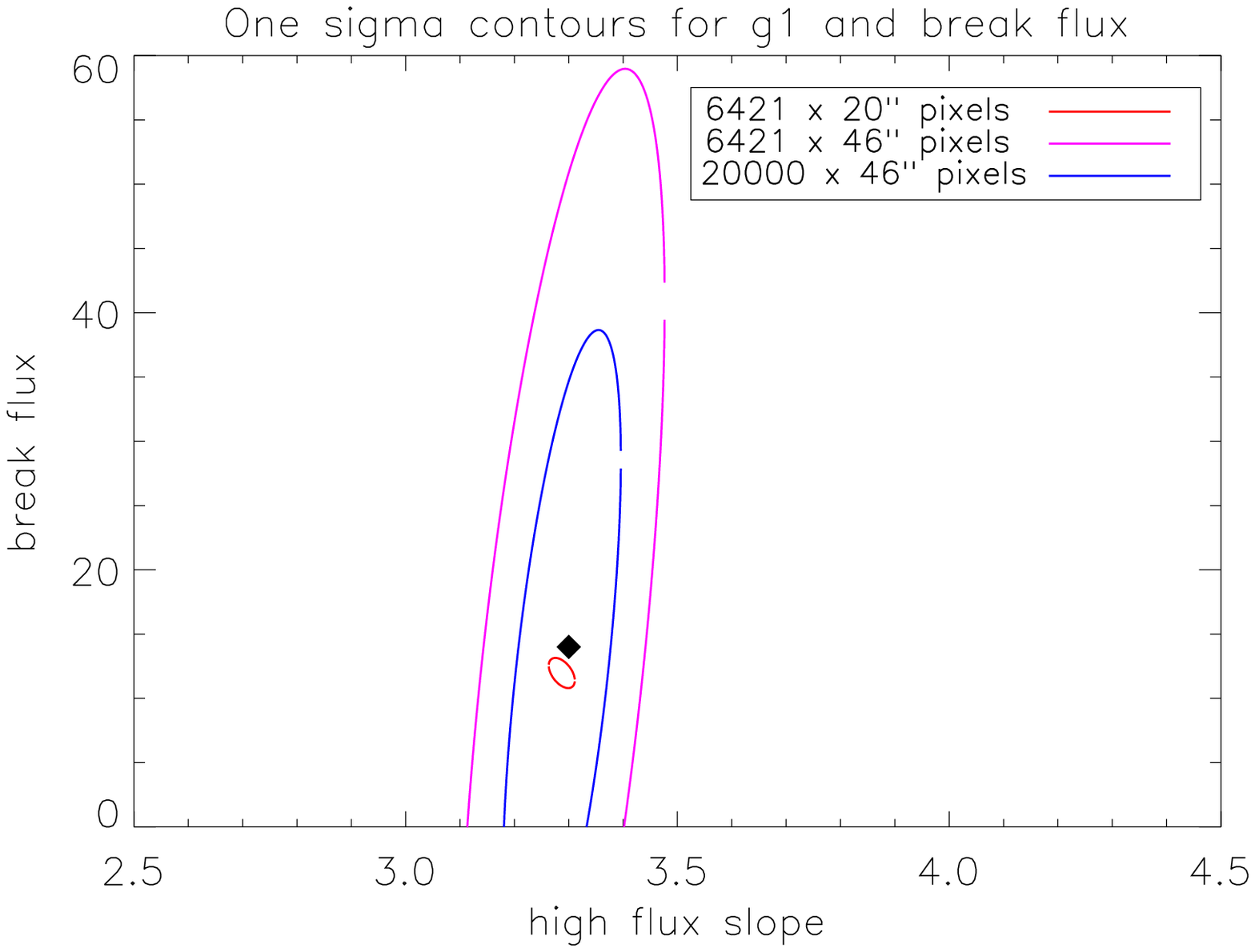} \\ 
\mbox{(a)} & \mbox{(b)}& \mbox{(c)}
\end{array}$
\end{center}
\caption{Here we present the 1$\sigma$ contours in the two dimensional
  plane of 3 different pairs of parameters, while marginalizing over
  the rest. These contours are calculated according to F$^{-1}$. The
  filled hexagons give the original parameters of the simulation on
  these planes.}
\label{ellipses}
\end{figure*}

Figure \ref{ellipses} also shows us that there are degeneracies
between the different parameters. These degeneracies  are not broken
when we use a more accurate experiment, they are ``built in''
through the definition of the model N(S). For example, as the model is
defined, the normalization is the number of sources with fluxes
greater than 1mJy. Naturally, if we increase $\gamma_1$ we should increase the 
normalization in order to remain within the error bars for N(S). 
In the following we will look further into the degeneracies by looking
at the derivatives of the model with respect to the parameters we are
interested in \cite[]{karh97}. The reason for this  can
be seen if we look again at the expression for the elements of the
Fisher Matrix (eq.\ref{fm}): they have the structure of a dot product between vectors
$\frac{\partial \mu _k}{\partial\theta_i}$ and
$\frac{\partial\mu _k}{\partial\theta_j}$ . If one of these vectors is 
a linear combination of another, the Fisher matrix elements will be singular, and the
errors of the estimated parameters will be infinite. If the vectors
are completely orthogonal then F and F$^{-1}$ will be diagonal and
thus there will be no correlation between different parameters and
their errors. Usually there will be some level of correlation which
will be manifested by a somewhat similar shape between the functional shape of
the different derivatives. By looking at the functional shape
of the derivatives one may recognize such degeneracies.
The way to avoid any degeneracies between parameters will be to
diagonalize the Fisher Matrix, thus changing the parameters into new
ones which are linear combinations of the originals. These parameters,
however, are not always physical, and thus do not have much meaning
unless they are very similar to the old ones, with not much
`contamination' from other parameters.

\begin{figure*}
\begin{center}
$\begin{array}{c@{\hspace{0.1in}}c}
\includegraphics[width=7cm]{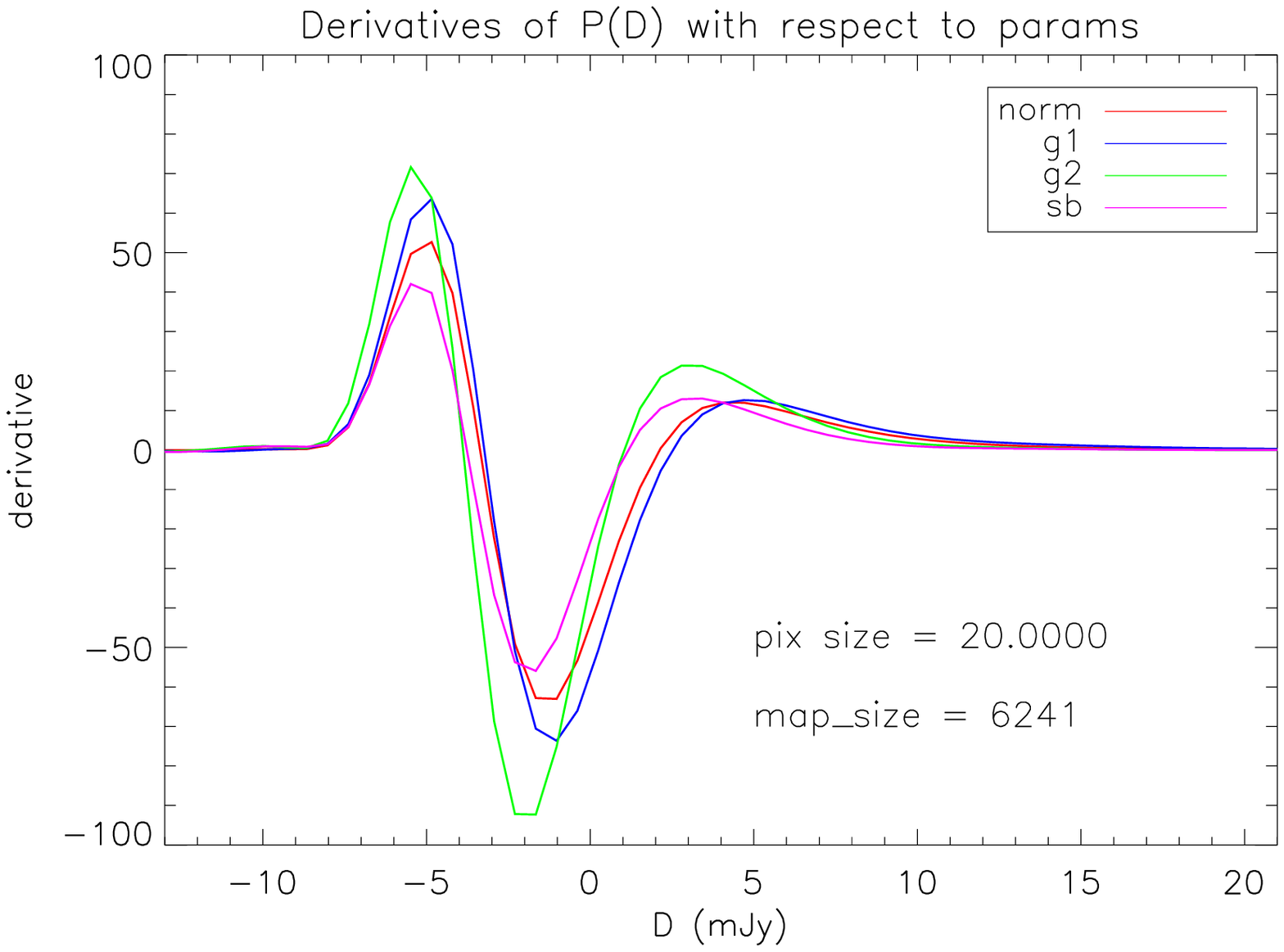}  &
\includegraphics[width=7cm]{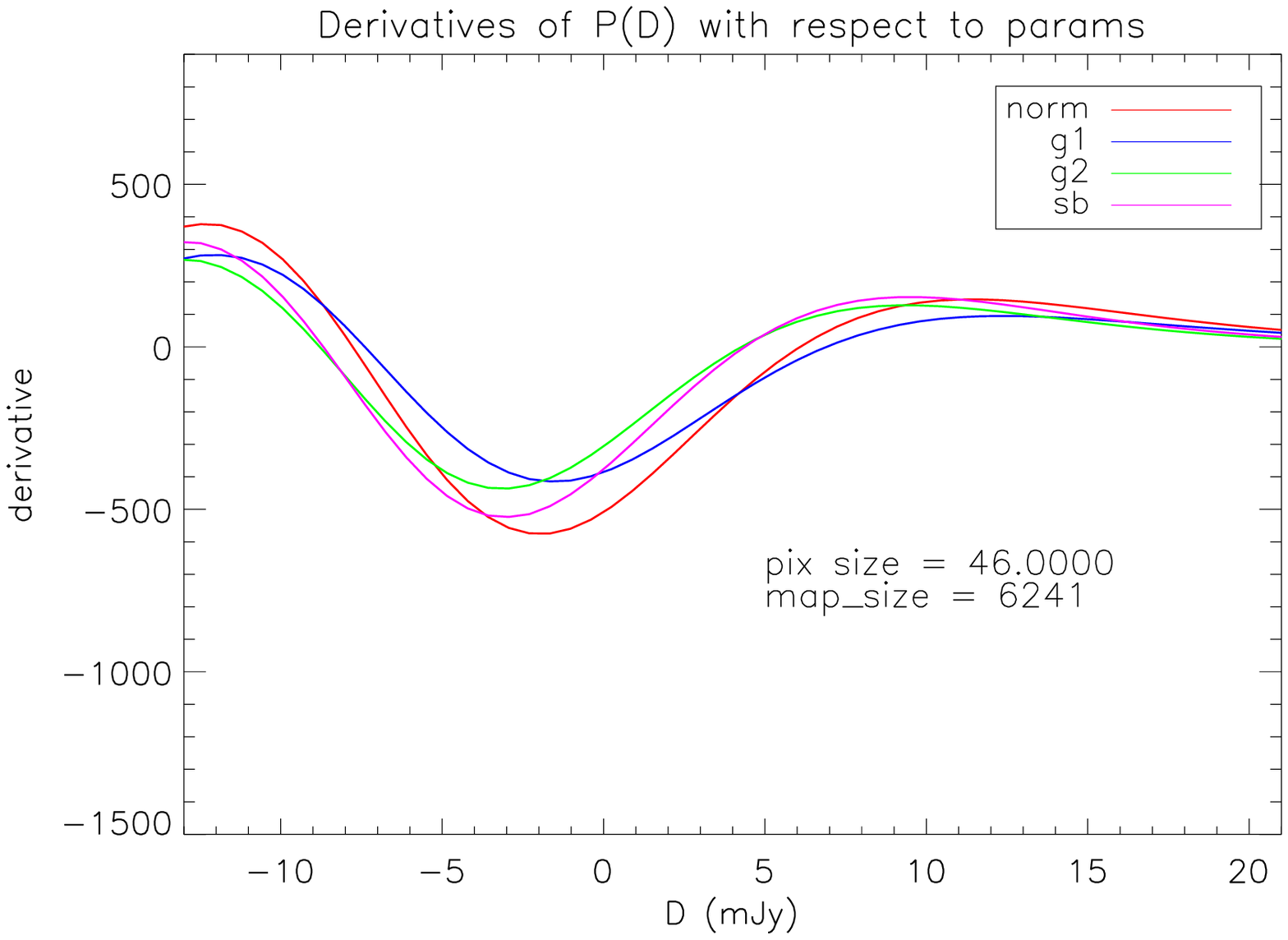} \\ 
\includegraphics[width=7cm]{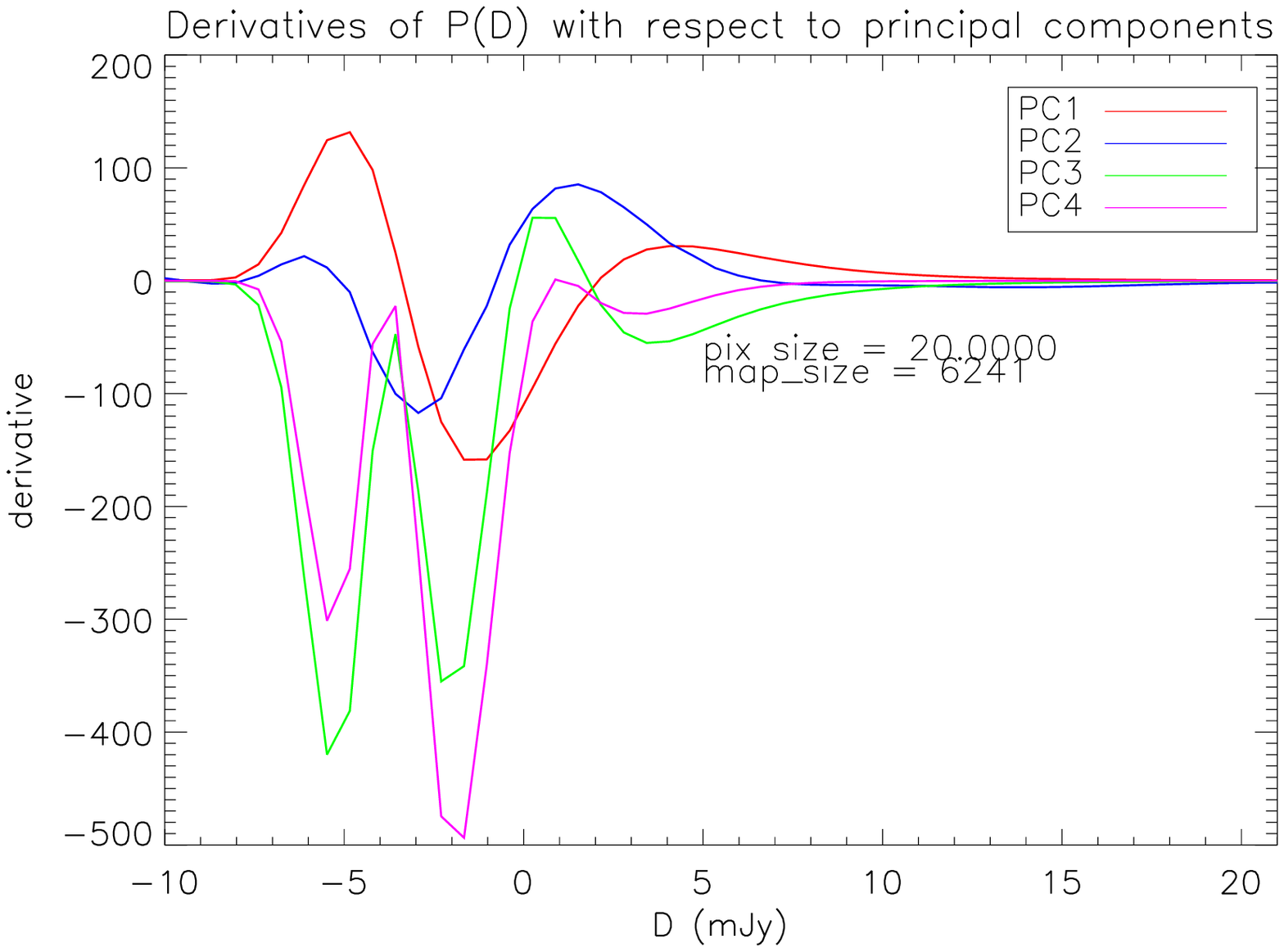}  &
\includegraphics[width=7cm]{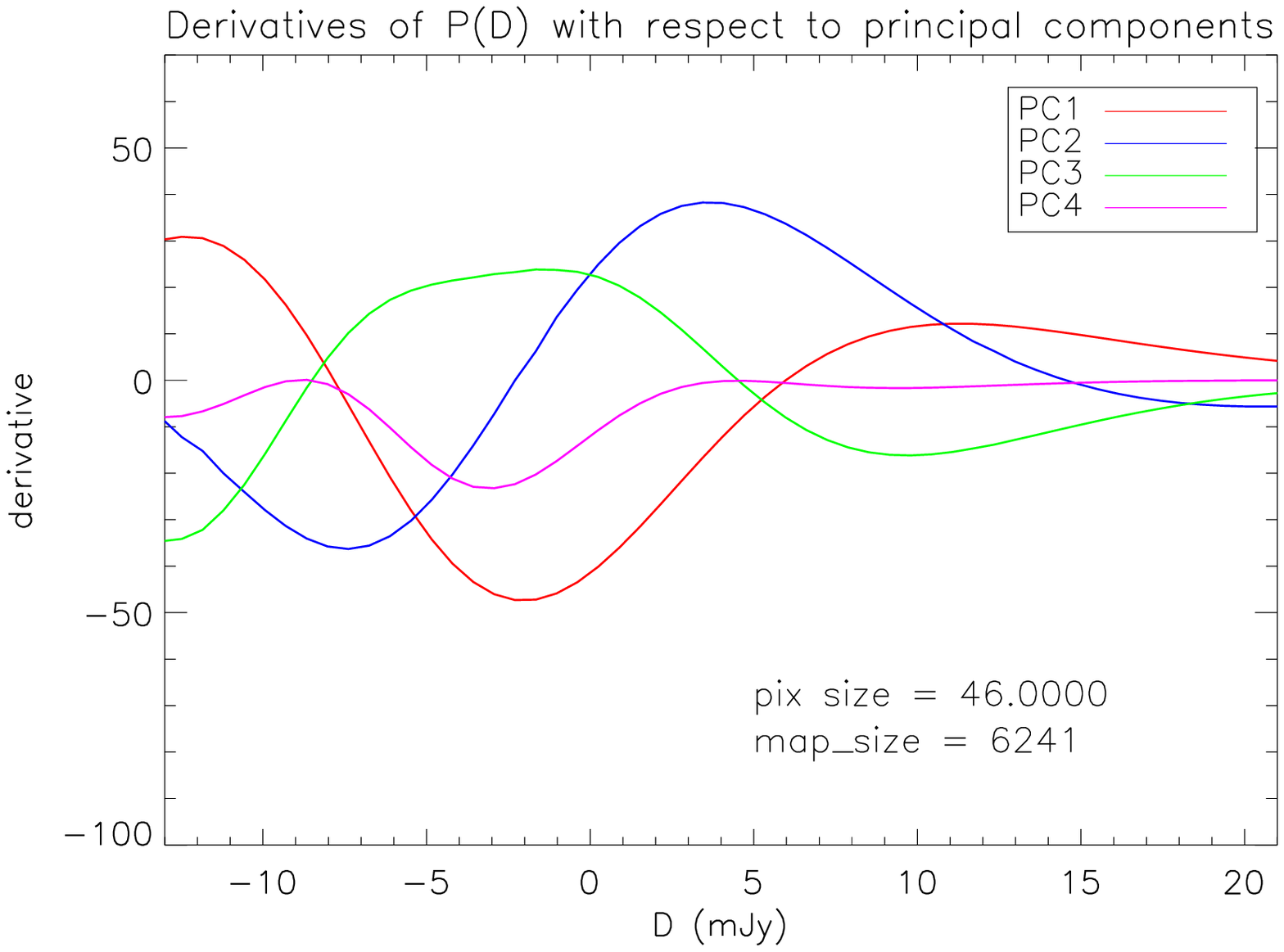} \\ 
\mbox{(a)} & \mbox{(b)}
\end{array}$
\end{center}
\caption{The top figures show the derivatives of the model with
  respect to the 4 parameters at 2 pixel sizes - 46$^2$'' and 20$^2$''.
the bottom show the derivatives of the model with respect to the
principal components. It is  possible to see how the degeneracies disappear.}
\label{degen}
\end{figure*}

It is worthwhile to check whether under improved experimental
conditions these degeneracies, if they exist, are removed or not.  
In fig \ref{degen} we plot the derivatives with respect to the usual
4 parameters and the derivatives with respect to the new,
``diagonalised'' parameters (called principal components - PC1-4). On the left 
panel we have the case for a pixel of 20*20 arcsec$^2$, and on the
right, 46*46 arcsec$^2$.  Again we can see that the degeneracies are
not removed by improving the experiment. Only diagonalizing the fisher
matrix allows us to remove them. 
We also add a table which specifies how the new parameters are built
from the old ones. In the first row we have the 4 parameters and in
the first column we have the 4 principal components. In each of the 4
rows we have the  coefficients of the linear combination. The
strongest coefficients are marked in bold. 
$$
\bordermatrix{&K &\gamma_1 &\gamma_2 &S_{break} \cr
PC1&  0.32   &   {\bf 0.91}  &   0.24  &  -0.03   \cr 
PC2&  {\bf 0.94}   &  -0.34  &   0.03  & -0.004    \cr 
PC3& -0.11   &  -0.21  &   {\bf 0.96}  &  -0.09    \cr 
PC4&  0.0015 &   0.002 &   0.098 &  {\bf 0.995}   } $$
PC1 has the highest weight from $\gamma_1$ with some contribution from
$K$ and $\gamma_2$. PC2 has the highest contribution from $K$ and some
contribution from $\gamma_1$. PC3 is mostly $\gamma_2$ and some
$\gamma_1$, and PC4 is almost exclusively $S_{break}$. The
degeneracies are to be expected since as mentioned before, P(D)
analysis cannot give information at fluxes much below the one source
per beam flux level. In our simulation this level is of the order of
7mJy and therefore it is reasonable that for example $\gamma_2$, which
is the slope of the counts below a few mJy is degenerate with the
other parameters. \footnote{We thank the referee for pointing out this
  cause for the degeneracies.}

\section{Discussion}

In this work we have explored the extent to which one can use a P(D)
analysis to gather information from far infra red sky maps. These maps
are characterized by a very high level of confusion noise which arises
due to the relatively poor resolution power available at these wavelengths.

We created a simulated map of the sky with an underlying modeled
N(S). The model consisted of a two slope model with a high flux slope
greater than 3 and a shallower low-flux slope. It was chosen this way
following the finding of a steep slope of the number counts of
resolved objects in the FIRBACK maps and in agreement with predictions
of galaxy evolution models. The parameters of the model are the two
slopes, the break flux (where the slope changes) and the normalization.

We then created the histogram of the simulated map and used it to find the
best fit parameters of the N(S) model and their errors. After finding
the best fitted parameters, we used the Fisher matrix analysis to
calculate the minimal errors possible of these parameters in
experiments with different pixel sizes and in experiments with
different total number of pixels, including those of the FIRBACK.

We found that our algorithm gives fitted parameters with almost the
minimal errors possible theoretically. The underlying parameters of
the simulated map were within 1$\sigma$ of the best fitted ones
(except for the normalization).
This means that the tool we have constructed in order to find
number counts is quite reliable.  

The Fisher Matrix analysis shows that in an experiment with pixel size of
the order ~10 arcsec we will be able to find all parameters with
errors of only about a few percent. The situation of the FIRBACK
experiment is quite different: it is only the high flux slope which
can be found with a small error bar of $\approx$4$\%$. This is
somewhat better than what was found by individual source extraction from
the maps (error of $\approx$20$\%$). 

The advantage of the P(D) analysis that it is sensitive down to the flux for
which there is one source per beam - in the FIRBACK case this is around
few mJy, much below the detection limit which is 180mJy. Also, the
extraction of even the high flux slope is straightforward and does not
warrant an a-priory extraction of sources or other manipulation of the maps.

In order to be able to extract the other parameters of the number
counts, we will have to wait for the SIRTF experiment. In that
experiment, the pixel size is 16''*16'' and the number of different
points measured on the sky is up an order of magnitude larger then for
FIRBACK. In this case the precision of estimation is enhanced due to
both factors: smaller pixels and more of them.

\section*{Acknowledgments}
We would like to thank Jeremy Blaizot for coding the instrumental
effects of ISOPHOT and for many discussions. Y.F thanks
Michel Fioc, Stephane Colombi, Bruno Guiderdoni, Guilaine Lagache,
Herve Dole, Ofer Lahav, Roberto Trotta, Jean-Pierre Eckmann,
Andreas Malaspinas, Tsvi Piran, Avishai Dekel and Yehuda Hoffman for
helpfull discussions. Y.F was funded by a Chateaubriand fellowship, a
Marie Curie grant and a Marie Heim-Voegtlin grant.  

\appendix

\section[]{Probability Distribution of Background Fluctuations}

The shape of the curve describing the probability that a pixel will accumulate a certain
level of flux depends on a few factors. These are the point spread 
function of the telescope, the pixel size and shape, and the source number 
counts N(S). 

We are assuming some properties for the sources which make up the
background: that the sources are point-like, they are not clustered,
and once a source is ``found'' its flux is a random variable
distributed according to the number-count-flux relation, $N(S)$.

The integrated flux seen by a pixel pointing in direction ${\mbox{\bf n}}$ will be:
     
\begin{equation}
S_{pix}({\mbox{\bf n}})=\int d\Omega_{{\mbox{\bf n'}}}G({\mbox{\bf n}}-{\mbox{\bf n'}})S({\mbox{\bf n}}')
\end{equation}

$d\Omega_{{\mbox{\bf n}}}$ is the solid angle in direction 
${\mbox{\bf n}}$. G$({\mbox{\bf n}}-{\mbox{\bf n'}})$ is the 
beam profile - it is the convolution of the  
point spread function of the imaging instrument and the pixel
profile.  $S({\mbox{\bf n}})$ is the flux per unit solid angle coming from
direction ${\mbox{\bf n}}$.   
Now we can calculate the probability density of $S_{pix}$ (the
probability that a certain pixel will receive flux $S_{pix}$). In order
to do that one first calculates the characteristic function of
$S_{pix}$ which is given by
\begin{equation}
\Phi(\omega)=< e^{2\pi i \omega S_{pix}} >
\end{equation}

Then the probability distribution for $S_{pix}$ will simply be the
Fourier transform of $\Phi$. 

Say there are  K sources. Then the total flux received by a pixel
should be written as the following discrete sum:
\begin{equation}
S_{pix}(\mbox{\boldmath $n$})=\sum_{i=1}^kS_iG(\mbox{\boldmath $n$}-\mbox{\boldmath $n_i$})
\end{equation}
The probability of finding K sources in the sky, with fluxes
$S_1,...S_k$ and in directions ${\bf n_1}$,...${\bf n_k}$, is
\begin{equation}
F^{(k)}(\mbox {\boldmath $n_1$},S_1...\mbox{\boldmath $n_k$},S_k)=\frac{(4\pi\mu )^k}{k!}e^{-4\pi\mu }(4\pi)^{-k}\prod_{i=1}^{k}f(S_i)
\end{equation}
$\mu $ is the mean number of sources per unit solid angle, and $f(S)$
is the probability that a source has a flux in the range
$[S,S+dS]$. This is a product of the Poissonian probability to find K
sources while the mean number of sources is expected to be $\mu $ and the
probability to find a source with flux $S_i$ given the model N(S), all
this per steradian (hence the term $(4\pi)^{-K}$. The
normalization condition is that the sum of the probabilities to find
any number of sources in any direction is one:
\begin{equation}
\sum_{k=0}^{\infty}\int d\Omega_{n_k}\int_0^{\infty} dS_1...
\int d \Omega_{\mbox{\boldmath $n_k$}}\int_0^{\infty} dS_k
F^{(k)}(\mbox{\boldmath $n_1$},S_1;...\mbox{\boldmath $n_k$},S_k)=1
\end{equation}
Using the normalization condition and the form of $S_{pix}$ we may now
calculate the characteristic function, which becomes:
\begin{equation}
\Phi(\omega)=exp\left[A_{pix}\int_0^{\infty}dS\frac{dN}{dS}(S)\sum_i\left(e^{2\pi
      i\omega P_iS}-1\right)\right]
\end{equation}
$P_i$ is the pixelized PSF and $A_{pix}$ is the area of the pixel in
steradians. The integration over the angles is separate from that over
flux and gives $A_{pix}$. This is justified
in our case because although the PSF is quite extended, outside of the
pixel it has very  small values. 
So now we may write the expression for the P($S_{pix}$), as the Fourier
transform of the characteristic function: 

\begin{equation}
P(S_{pix})=\int_{-\infty}^{\infty}\Phi(\omega)e^{-2\pi i\omega S_{pix}}\omega
\end{equation}

The above expression gives the probability that a pixel will receive
an amount of flux equal to $S_{pix}$. We will be working with a slightly
different expression - the probability to get a certain flux above or
below the mean flux on the map, $D$, defined as $S_{pix}=D+<S_{pix}>$

\begin{equation}
P(D)=\int_{-\infty}^{\infty}\Phi(\omega)e^{-2\pi i\omega D}\omega
\end{equation}

Once we have P(D) we may integrate between the bins and multiply by
the total number of pixels in the map to get the expected number of
pixels falling within each bin.

\def\ref@jnl#1{{\rmfamily #1}}%

\newcommand\aj{\ref@jnl{AJ}}%


\newcommand\araa{\ref@jnl{ARA\&A}}%


\newcommand\apj{\ref@jnl{ApJ}}%


\newcommand\apjl{\ref@jnl{ApJ}}%
 

\newcommand\apjs{\ref@jnl{ApJS}}%


\newcommand\ao{\ref@jnl{Appl.~Opt.}}%


\newcommand\apss{\ref@jnl{Ap\&SS}}%


\newcommand\aap{\ref@jnl{A\&A}}%


\newcommand\aapr{\ref@jnl{A\&A~Rev.}}%


\newcommand\aaps{\ref@jnl{A\&AS}}%


\newcommand\azh{\ref@jnl{AZh}}%


\newcommand\baas{\ref@jnl{BAAS}}%


\newcommand\jrasc{\ref@jnl{JRASC}}%


\newcommand\memras{\ref@jnl{MmRAS}}%


\newcommand\mnras{\ref@jnl{MNRAS}}%


\newcommand\pra{\ref@jnl{Phys.~Rev.~A}}%


\newcommand\prb{\ref@jnl{Phys.~Rev.~B}}%


\newcommand\prc{\ref@jnl{Phys.~Rev.~C}}%


\newcommand\prd{\ref@jnl{Phys.~Rev.~D}}%


\newcommand\pre{\ref@jnl{Phys.~Rev.~E}}%


\newcommand\prl{\ref@jnl{Phys.~Rev.~Lett.}}%


\newcommand\pasp{\ref@jnl{PASP}}%


\newcommand\pasj{\ref@jnl{PASJ}}%


\newcommand\qjras{\ref@jnl{QJRAS}}%


\newcommand\skytel{\ref@jnl{S\&T}}%


\newcommand\solphys{\ref@jnl{Sol.~Phys.}}%


\newcommand\sovast{\ref@jnl{Soviet~Ast.}}%


\newcommand\ssr{\ref@jnl{Space~Sci.~Rev.}}%


\newcommand\zap{\ref@jnl{ZAp}}%


\newcommand\nat{\ref@jnl{Nature}}%


\newcommand\iaucirc{\ref@jnl{IAU~Circ.}}%


\newcommand\aplett{\ref@jnl{Astrophys.~Lett.}}%


\newcommand\apspr{\ref@jnl{Astrophys.~Space~Phys.~Res.}}%


\newcommand\bain{\ref@jnl{Bull.~Astron.~Inst.~Netherlands}}%


\newcommand\fcp{\ref@jnl{Fund.~Cosmic~Phys.}}%


\newcommand\gca{\ref@jnl{Geochim.~Cosmochim.~Acta}}%


\newcommand\grl{\ref@jnl{Geophys.~Res.~Lett.}}%


\newcommand\jcp{\ref@jnl{J.~Chem.~Phys.}}%


\newcommand\jgr{\ref@jnl{J.~Geophys.~Res.}}%


\newcommand\jqsrt{\ref@jnl{J.~Quant.~Spec.~Radiat.~Transf.}}%


\newcommand\memsai{\ref@jnl{Mem.~Soc.~Astron.~Italiana}}%


\newcommand\nphysa{\ref@jnl{Nucl.~Phys.~A}}%


\newcommand\physrep{\ref@jnl{Phys.~Rep.}}%


\newcommand\physscr{\ref@jnl{Phys.~Scr}}%


\newcommand\planss{\ref@jnl{Planet.~Space~Sci.}}%


\newcommand\procspie{\ref@jnl{Proc.~SPIE}}%


\let\astap=\aap 

\let\apjlett=\apjl 

\let\apjsupp=\apjs 

\let\applopt=\ao


\begin{thebibliography}{99}

\bibitem[{Barcons}(1992)]{barcons92}
{Barcons}, X. (1992).
\newblock {Confusion noise and source clustering}.
\newblock {\em \apj}, {\bf 396}, 460--468.

\bibitem[{Barcons} \& {Fabian}(1990)]{bar&fab90}
{Barcons}, X. \& {Fabian}, A.~C. (1990).
\newblock {A fluctuation analysis of the X-ray background in the Einstein
  Observatory IPC }.
\newblock {\em \mnras}, {\bf 243}, 366--371.

\bibitem[{Condon}(1974)]{condon74}
{Condon}, J.~J. (1974).
\newblock {Confusion and Flux-Density Error Distributions}.
\newblock {\em \apj}, {\bf 188}, 279--286.

\bibitem[{Dole} et al.(2001)]{firback3}
{Dole}, H., {Gispert}, R., {Lagache}, G., {Puget}, J.-L., {Bouchet}, F.~R.,
  {Cesarsky}, C., {Ciliegi}, P., {Clements}, D.~L., {Dennefeld}, M., {D{\'
  e}sert}, F.-X., {Elbaz}, D., {Franceschini}, A., {Guiderdoni}, B., {Harwit},
  M., {Lemke}, D., {Moorwood}, A.~F.~M., {Oliver}, S., {Reach}, W.~T.,
  {Rowan-Robinson}, M., \& {Stickel}, M. (2001).
\newblock {FIRBACK: III. Catalog, source counts, and cosmological implications
  of the 170 mu m ISO}.
\newblock {\em \aap}, {\bf 372}, 364--376.

\bibitem[{Dole} et al.(2003)]{dole_sirtf}
{Dole}, H., {Lagache}, G., \& {Puget}, J.-L. (2003).
\newblock {Predictions for Cosmological Infrared Surveys from Space with the
  Multiband Imaging Photometer for SIRTF}.
\newblock {\em \apj}, {\bf 585}, 617--629.

\bibitem[{Efstathiou}(1999)]{efst99}
{Efstathiou}, G. (1999).
\newblock {Constraining the equation of state of the Universe from distant Type
  Ia supernovae and cosmic microwave background anisotropies}.
\newblock {\em \mnras}, {\bf 310}, 842--850.

\bibitem[{Guiderdoni} et al.(1998)]{guider98}
{Guiderdoni}, B., {Hivon}, E., {Bouchet}, F.~R., \& {Maffei}, B. (1998).
\newblock {Semi-analytic modelling of galaxy evolution in the IR/submm range}.
\newblock {\em \mnras}, {\bf 295}, 877--898.

\bibitem[{Haiman} \& {Knox}(2000)]{haiman00}
{Haiman}, Z. \& {Knox}, L. (2000).
\newblock {Correlations in the Far-Infrared Background}.
\newblock {\em \apj}, {\bf 530}, 124--132.

\bibitem[{Jungman} et al.(1996)]{weigh96}
{Jungman}, G., {Kamionkowski}, M., {Kosowsky}, A., \& {Spergel}, D.~N. (1996).
\newblock {Weighing the universe with the cosmic microwave background}.
\newblock {\em Physical Review Letters}, {\bf 76}, 1007--1010.

\bibitem[{Knox} et al.(2001)]{knox01}
{Knox}, L., {Cooray}, A., {Eisenstein}, D., \& {Haiman}, Z. (2001).
\newblock {Probing Early Structure Formation with Far-Infrared Background
  Correlations}.
\newblock {\em \apj}, {\bf 550}, 7--20.

\bibitem[{Lagache} \& {Dole}(2001)]{firback2}
{Lagache}, G. \& {Dole}, H. (2001).
\newblock {FIRBACK. II. Data reduction and calibration of the 170 $\mu$ m ISO
  deep cosmological survey}.
\newblock {\em \aap}, {\bf 372}, 702--709.

\bibitem[{Lagache} \& {Puget}(2000)]{lag&puget00a}
{Lagache}, G. \& {Puget}, J.~L. (2000).
\newblock {Detection of the extra-Galactic background fluctuations at 170 mu
  m}.
\newblock {\em \aap}, {\bf 355}, 17--22.

\bibitem[{Lagache} et al.(2000)]{lag&pug00}
{Lagache}, G., {Puget}, J., {Abergel}, A., {Desert}, F., {Dole}, H., {Bouchet},
  F.~R., {Boulanger}, F., {Ciliegi}, P., {Clements}, D.~L., {Cesarsky}, C.,
  {Elbaz}, D., {Franceschini}, A., {Gispert}, R., {Guiderdoni}, B., {Haffner},
  L.~M., {Harwit}, M., {Laureijs}, R., {Lemke}, D., {Moorwood}, A.~F.~M.,
  {Oliver}, S., {Reach}, W.~T., {Reynolds}, R.~J., {Rowan-Robinson}, M.,
  {Stickel}, M., \& {Tufte}, S.~L. (2000).
\newblock {The Extragalactic Background and Its Fluctuations in the
  Far-Infrared Wavelengths}.
\newblock In {\em ISO Survey of a Dusty Universe, Proceedings of a Ringberg
  Workshop Held at Ringberg Castle, Tegernsee, Germany, 8-12 November 1999,
  Edited by D. Lemke, M. Stickel, and K. Wilke, Lecture Notes in Physics, vol.
  548, p.81}, pages 81--+.

\bibitem[{Puget} et al.(1999)]{firback1}
{Puget}, J.~L., {Lagache}, G., {Clements}, D.~L., {Reach}, W.~T., {Aussel}, H.,
  {Bouchet}, F.~R., {Cesarsky}, C., {D{\' e}sert}, F.~X., {Dole}, H., {Elbaz},
  D., {Franceschini}, A., {Guiderdoni}, B., \& {Moorwood}, A.~F.~M. (1999).
\newblock {FIRBACK. I. A deep survey at 175 microns with ISO, preliminary
  results}.
\newblock {\em \aap}, {\bf 345}, 29--35.

\bibitem[{Scheuer}(1974)]{sch74}
{Scheuer}, P.~A.~G. (1974).
\newblock {Fluctuations in the X-ray background}.
\newblock {\em \mnras}, {\bf 166}, 329--338.

\bibitem[{Scott} \& {White}(1999)]{scott99}
{Scott}, D. \& {White}, M. (1999).
\newblock {Implications of SCUBA observations for the Planck Surveyor}.
\newblock {\em \aap}, {\bf 346}, 1--6.

\bibitem[{Takeuchi} et al.(2001)]{takeuchi}
{Takeuchi}, T.~T., {Ishii}, T.~T., {Hirashita}, H., {Yoshikawa}, K.,
  {Matsuhara}, H., {Kawara}, K., \& {Okuda}, H. (2001).
\newblock {Exploring Galaxy Evolution from Infrared Number Counts and Cosmic
  Infrared Background}.
\newblock {\em \pasj}, {\bf 53}, 37--52.

\bibitem[{Tegmark} et al.(1997)]{karh97}
{Tegmark}, M., {Taylor}, A.~N., \& {Heavens}, A.~F. (1997).
\newblock {Karhunen-Loeve Eigenvalue Problems in Cosmology: How Should We
  Tackle Large Data Sets?}
\newblock {\em \apj}, {\bf 480}, 22--+.

\end{thebibliography}

\end{document}